# Pressure-driven phase transformations and phase segregation in ferrielectric CuInP$_2$S$_6$-In$_{4/3}$P$_2$S$_6$ self-assembled heterostructures

Rahul Rao,[1*] Benjamin S. Conner,[2,3] Ryan Selhorst,[1,4] Michael A. Susner[1]

[1]Materials and Manufacturing Directorate, Air Force Research Laboratory, 2179 12th Street, Wright-Patterson Air Force Base, Ohio 45433, USA

[2]Sensors Directorate, Air Force Research Laboratory, 2241 Avionics Circle, Wright-Patterson Air Force Base, Ohio 45433, USA

[3]National Research Council, Washington, D.C. 20001, USA

[4]UES Inc., Dayton, Ohio 45433, USA

**Abstract**

Layered multi-ferroic materials exhibit a variety of functional properties that can be tuned by varying the temperature and pressure. As-synthesized CuInP$_2$S$_6$ is a layered material that displays ferrielectric behavior at room temperature. When synthesized with Cu deficiencies, CuInP$_2$S$_6$ spontaneously phase segregates to form ferrielectric CuInP$_2$S$_6$ (CIPS) and paraelectric In$_{4/3}$P$_2$S$_6$ (IPS) domains in a two-dimensional self-assembled heterostructure. Here, we study the effect of hydrostatic pressure on the structure of Cu-deficient CuInP$_2$S$_6$ by Raman spectroscopy measurements up to 20 GPa. Detailed analysis of the frequencies, intensities, and linewidths of the Raman peaks reveals four discontinuities in the spectra around 2, 10, 13 and 17 GPa. At ~2 GPa, we observe a structural transition initiated by the diffusion of IPS domains, which culminates in a drastic reduction of the number of peaks around 10 GPa. We attribute this to a possible monoclinic-trigonal phase transition at 10 GPa. At higher pressures (~ 13 GPa), significant increases in peak intensities and sharpening of the Raman peaks suggest a bandgap-lowering and an isostructural electronic transition, with a possible onset of metallization at pressures above 17 GPa. When the pressure is released, the structure again phase-separates into two distinct chemical domains within the same single crystalline framework- however, these domains are much smaller in size than the as-synthesized material resulting in suppression of ferroelectricity through nanoconfinement. Hydrostatic pressure can thus be used to tune the electronic and ferrielectric properties of Cu-deficient layered $CuInP_2S_6$.

[*] Correspondence – rahul.rao.2@us.af.mil

1. **Introduction**

Two-dimensional (2D) materials have recently come into focus due to their small form factors which enable unique thermal, electrical, and magnetic properties of interest for next-generation devices [1–4]. Various materials have been investigated for the purposes of exploring and manipulating functional (i.e. magnetism and ferroelectricity) phase-spaces at the nanoscale [5]. Metal thio- and seleno-phosphates (collectively abbreviated here as MTPs) have proven to be valuable in this respect as this materials family intrinsically exhibits the functionalities needed for next-generation microelectronic devices [6,7]. Their structural backbone is comprised of $[P_2S_6]^{4-}$ or $[P_2Se_6]^{4-}$ ethane-like anion groups that ionically pair with hexagonally arranged metal cations. The magnetic moments, or the preferred positions of the metal cations or inside the octahedral cage formed by the chalcogen atoms (leading to electric polarization), are directly responsible for the ferroic ordering noted in many MTP compositions [6–8].

$CuInP_2S_6$ is a van der Waals layered material of interest due to its ferrielectric nature and above-room temperature Curie temperature ($T_C$ ~ 315 K in bulk $CuInP_2S_6$). The nature of the ferroelectric transition in this material is due to the preferred positions of the Cu cations within an individual lamella. Above 315 K, the Cu atoms equally occupy both the up and down positions within its octahedron; below 315 K Cu atoms prefer to be in the up position, leading to spontaneous electric polarization [9].

Recent work has shown that, when synthesized with Cu deficiencies, the material spontaneously phase separates into $CuInP_2S_6$ (CIPS) and $In_{4/3}P_2S_6$ (IPS) domains within the same single crystal with a common $[P_2S_6]^{4-}$ anionic foundation [10,11]; the overall composition can be described as the charge-balanced $Cu_{1-x}In_{1+x/3}P_2S_6$. Interestingly, these domains disappear upon heating above 500 K when the cations simply "melt" and form a mobile cation liquid inside the rigid anion lattice [11]. When cooled below 500 K, the two phases reform; the size of these domains (and at a microscopic level, the confinement of ferroelectric ordering [11,12]) depends on the kinetics through the transition. Rapid quenching results in nanoscale CIPS domains and suppressed ferroelectric ordering. The interplay between the two different crystal structures leads to an increase in the overall $T_C$ of the ferroelectric CIPS phase [10,11] as well as the presence of a quadruple potential well for the Cu atomic ordering [13] that appears to be largely tunable.

Although some progress has been made with respect to understanding the relationships between temperature, structure, and ferroelectric ordering, very little has been done with respect to investigating the effects of pressure outside phenomenological studies on the pure phase CIPS that show increases in $T_C$ with pressure [14–17]. However, no correlation with structure has been proposed. Spurred by this and recent observations of pressure-induced phase transitions in a variety of MTP materials such as $CuInP_2S_6$ [14], $SnP_2S_6$ [18], $NiPS_3$ [19], $CrPS_4$ [20], $FePS_3$ [21], $MnPS_3$ [22] as well as other 2D materials [23], we investigate the effects of hydrostatic pressure on structural ordering and phase transitions in $CuInP_2S_6$-$In_{4/3}P_2S_6$ (CIPS-IPS) self-assembled heterostructures using Raman spectroscopy.

We measure Raman spectra from CIPS-IPS single crystals under hydrostatic compression up to 20 GPa. Analysis of the spectral data, combined with temperature-dependent Raman spectra and electron microscopy reveal discontinuities at four pressures, which we attribute to transitions in the CIPS-IPS heterostructures. We observe the initiation of structural disorder through the distortion of the IPS sub-lattice around 2 GPa, followed by a monoclinic-to-trigonal transition around 10 GPa. At the highest pressure, we observe anomalous sharpening of Raman modes and a decrease in peak intensities, which we tentatively attribute to electronic transitions and metallization. Finally, our data also suggest phase segregation of $CuInP_2S_6$ and $In_{4/3}P_2S_6$ into nanoscale domains, similar to what has previously been shown for rapid quenching upon cooling from above 500 K to room temperature.

## 2. Experimental Methods

2.1 Synthesis, Chemical, and Structural Characterization

The details of the crystal synthesis have been described elsewhere [10]. Briefly, single crystals of CIPS-IPS are synthesized through standard solid state synthesis techniques. The precursor $In_2S_3$ (prepared from Alfa Aesar Puratronic elements, 99.999% purity, sealed in an evacuated fused silica ampoule and reacted at 950 °C for 48 hours) is reacted with the necessary quantities of Cu (Alfa Aesar Puratronic ), P (Alfa Aesar Puratronic ) and S (Alfa Aesar Puratronic ) to obtain the composition $Cu_{0.4}In_{1.2}P_2S_6$. The starting materials are sealed in a fused silica ampoule with ~80 mg of $I_2$ and loaded into a tube furnace. The furnace is slowly ramped to 775°C over a period of 24 hours and held at that temperature for 100 hours. Afterwards, the sample is cooled at a rate of 20°C/hr. This latter step is crucial in forming large domains as their size largely depends on kinetics [24]. The total composition, both with and without compression, is confirmed via electron dispersion spectroscopy (EDS) analysis in conjunction with a Zeiss Gemini scanning electron microscope.

High-resolution synchrotron data are obtained on beamline 11-ID-C at the Advanced Photon Source at Argonne National Laboratory. Temperature-dependent X-ray diffraction (XRD) patterns are collected in a Linkam THM600 heating/cooling stage. The samples are ground into a fine powder and carefully placed in a Cu cylinder to avoid issues related to preferential orientation in this layered material. This cylinder is in turn placed in the Linkam stage and purged with inert gas. XRD patterns are taken from low temperatures to high (with $\Delta T < 0.2$ K) as the CIPS-IPS heterostructure is sensitive to thermal history. XRD patterns are collected in the transmission mode using a Perkin-Elmer large area detector ( with a synchrotron radiation wavelength of 0.117418 Å). The collected 2D patterns are processed into conventional 1D patterns (Fit2d software) and further refined using the FullProf suite [25].

2.2 Optical characterization

The pressure studies are conducted in a Merrill-Bassett type diamond anvil cell (DAC) with Boehler-Almax anvils. The DAC is fitted with Type II-as diamonds with 500 µm culets, affording a maximum pressure of 20 GPa. A stainless steel gasket with a 90 µm aperture is placed between the anvils, and a CIPS-IPS crystal is loaded into the cell along with a small ruby crystal as a standard. The chamber is then filled with a 4:1 ratio methanol/ethanol solution which acts as the pressure transmitting medium [26]. The pressure is measured using the R1 fluorescence peak of the ruby [27]. Room temperature pressure-dependent Raman spectra are collected using a Renishaw inVia Raman microscope, with a 632.8 nm excitation laser focused through a 50x magnification long-working distance objective lens on to the CIPS-IPS crystal within the DAC. Screw-driven pressure is exerted on the sample manually and spectra are collected with an excitation laser power of ~1 µW to minimize heating. Additional temperature-dependent Raman spectra (up to 350 K, just above paraelectric phase transition temperature ~ 340 K) are collected in a Linkam hot/cold microscope stage under an argon atmosphere. Spectral analysis is performed (in Igor Pro) by normalizing intensities with respect to the diamond Raman peak, followed a spline baseline subtraction and Lorentzian peak fitting to extract frequencies and linewidths.

2.3 Morphological Characterization

Atomic Force Microscopy (AFM) is performed on a Bruker Dimension Icon AFM equipped with Nanoscope V controller. All measurements are performed under ambient conditions using tapping mode. TESPA (Nanoworld) AFM tapping mode tips are used with a force constant of 42 N/m and a resonance frequency of 320 kHz. Unpressurized CIPS-IPS crystals are cleaved and placed onto a Si/SiO$_2$ substrate. Freshly cleaved flat surfaces are chosen for imaging. The de-pressurized CIPS-IPS samples are scanned directly, within the 90 μm aperture, in the stainless steel gasket used for the Raman pressure study described above.

3. Results and Discussion

Schematics of the parent phase CuInP$_2$S$_6$ (CIPS) and In$_{4/3}$P$_2$S$_6$ (IPS) crystal structures are shown in Fig. 1a and 1b, respectively. Under ambient conditions (ferrielectric phase), the CIPS crystal is monoclinic (space group *Cc*), and consists of S$_6$ octahedra bridged by P-P pairs. The metal cations occupy off-center sites within these octahedra; their positions are attributed to a second-order Jahn-Teller instability associated with the d$^{10}$ electronic configuration of Cu [28]. The antiparallel displacements of the Cu$^{1+}$ and In$^{3+}$ ions account for the room temperature ferrielectric ordering in CIPS with polarization perpendicular to the layers. IPS, on the other hand, is paraelectric (also monoclinic, space group *P2/c*) at room temperature and its structure consists of an ordered arrangement of In$^{3+}$ ions and vacant sites within the octahedral network.

When synthesized in Cu-deficient conditions, the resultant single crystals exhibit interesting chemical behavior. Rather than the formation of an alloy, the crystals are comprised of two distinct domains - CuInP$_2$S$_6$ and In$_{4/3}$P2S$_6$ - in a ratio the [P$_2$S$_6$]$^{4-}$ backbone common between these two crystal structures allows for the formation of distinct chemical domains within the crystal. At $T$ > 500K, these chemical domains vanish and the cations form a highly mobile melt within the rigid anion backbone. The size of these domains depends on the cooling rate from the uniform high-temperature phase [10,11]. Structural analysis of a representative composition of the as grown, slow-cooled Cu-deficient phase (Composition Cu$_{0.4}$In$_{1.2}$P$_2$S$_6$) is enabled through refinement of synchrotron XRD data (Fig. 1c). The presence of Cu in trigonal positions at or near the outer boundary of the lamellae in CuInP$_2$S$_6$ serves to increase layer spacing (the sum of one lamella and one van der Waals gap) compared to that of In$_{4/3}$P$_2$S$_6$ (where the cations are firmly ensconced at the lamella midplane. In addition, the presence of Cu in CIPS, along with vacant sites in IPS serves to increase the area per P$_2$S$_6$ structural unit [29] in the Cu-containing compound. Thus, the unit cell volume is larger in the native CIPS phase. When CIPS and IPS are stitched together in a heterostructure, the chemical pressure exerted on the CIPS by the smaller IPS increases the $T_C$ of the ferroelectric transition, following the Clausius-Clapeyron relationship [10]. This pressure also results in a strain exerted by the CIPS phase on the IPS domains when the two are in a heterostructured form. We can see this in the diffraction data where the strains associated with the ferrielectric transition in heterostructured CIPS are carried over to the IPS when the two are in intimate contact (Fig. 1c). This apparent transition in the otherwise paraelectric IPS has consequences in the Raman data analyzed below.

To thoroughly characterize the ferroelectric transition in CIPS-IPS and how it differs from the pure-phase CIPS and IPS materials, we next compare the Raman spectra of these materials. An unpolarized, room temperature, ambient pressure Raman spectrum (collected with 633 nm laser excitation) from CIPS is shown in Fig. 2 (bottom trace). The spectrum exhibits several peaks between 50 – 650 cm$^{-1}$ that consist of external (<150 cm$^{-1}$) and internal (>150 cm$^{-1}$) vibrations. The vibrational modes can be divided into five frequency ranges, and in general are common to all layered MTPs. The lowest frequency peaks (< 50 cm$^{-1}$) and those between 50 and 150 cm$^{-1}$ correspond to cation and anion librations, respectively. The peaks between 150 – 200 cm$^{-1}$ and 200 – 350 cm$^{-1}$ correspond to deformations of the S-P-P and S-P-S bonds within the octahedra, respectively (or collectively, anion deformation modes). The high intensity peak around 370 cm$^{-1}$ is attributed to P-P vibrations, and the P-S (anion) stretching vibrations appear around 450 and 550 cm$^{-1}$. Within these groups, some anion deformation modes are sensitive to the type of metal cation. In particular, the peak around 320 cm$^{-1}$ can be attributed to distortions within the S$_6$ cage occupied by the Cu$^+$ ions [30]. This is likewise also applicable for the high frequency stretching modes where the lower (higher) frequency modes are influenced by Cu (In) cations [31,32].

The spectrum from CIPS-IPS (top trace in Fig. 2) also exhibits peaks grouped in the same frequency ranges described above, but is much more complex than the CIPS spectrum. It contains more peaks that can be attributed to the CIPS and IPS sub-lattices within the material. In all, we resolve 27 peaks in the room temperature Raman spectrum (fitted spectrum is shown in Fig. S1 in the Supplemental Information [33]), and the peak frequencies along with their vibrational mode assignments are listed in Table 1. A comparison of the CIPS-IPS Raman spectrum with spectra from the pure phase CIPS (bottom trace, Fig. 2) and IPS (Ref. [34]) enables us to identify peaks unique to both sub-lattices – there are four peaks in the anion libration region between 100 and 140 cm$^{-1}$, two of which can be attributed to CIPS (100 and 114 cm$^{-1}$) and the other two to IPS (127 and 140 cm$^{-1}$). Similarly, between 200 and 300 cm$^{-1}$, the two most intense peaks appear at ~255 and ~270 cm$^{-1}$, and can be assigned to IPS and CIPS, respectively. A peak at 300 cm$^{-1}$ in the CIPS-IPS spectrum does not appear in the CIPS spectrum, thus it can be attributed to anion deformation in IPS. Furthermore, as previously mentioned, the peak at 320 cm$^{-1}$ can be assigned to anion deformation in CIPS. Among the anion stretching modes, the two highest frequency peaks around 580 and 610 cm$^{-1}$ do not appear in the CIPS spectrum and therefore can be attributed to IPS. We note that while we have compared the CIPS-IPS Raman spectrum to pure-phase CIPS and IPS, we cannot preclude mode mixing or frequency shifts due to strains within the CIPS and IPS sub-lattices within the CIPS-IPS lamellae.

Next, we discuss the effect of hydrostatic pressure on the phonon modes in CIPS-IPS. Figs. 3a and 3b show Raman spectra upon compression (up to 20 GPa) and decompression, respectively. The spectra in Fig. 3 are normalized with respect to the highest peak intensity within the plotted range and vertically offset for clarity. A few general observations can be made at the outset – the spectra undergo changes in peak intensities and an overall blueshift in peak frequencies upon compression. In addition, the Raman spectrum fully recovers upon decompression. While we do not know if the layered structures persist under compression to high pressures, the recovery confirms that the layered structure can be regained upon decompression, without any significant loss of structure. Beyond these general observations, several subtle changes upon compression are observable in the spectra and described in detail below.

Going up in pressure from the ambient to 2.5 GPa, we observe an abrupt reduction in intensities of several peaks between 50 and 350 cm$^{-1}$ (bottom two spectra in Fig. 3a). A previous temperature-dependent Raman study [30] has shown a significant modulation of intensities of the low frequency peaks when pure phase CIPS is heated above its ferrielectric-paraelectric transition (~ 315 K). It is therefore tempting to ascribe our pressure-induced intensity modulations to a ferrielectric-paraelectric transition. However, the $T_C$ in CIPS-IPS (~340 K [10,11]) is higher than the pure phase CIPS ($T_C$ ~ 315 K). Moreover, the $T_C$ increases linearly with hydrostatic pressure compression [17,35]. A higher symmetry phase above 5 GPa has been observed in CIPS [14], although it is unclear if ferroic ordering is present at high pressures. Considering that our measurements were performed at 298 K (~40 K below the enhanced $T_C$ of the CIPS-IPS heterostructure), it is therefore unlikely that we are able to observe a ferrielectric-paraelectric transition upon compression.

Nevertheless, it is instructive to compare the pressure-dependent Raman spectra to temperature-dependent spectra collected just above $T_C$. Fig. 4 shows this comparison, with the Raman spectrum from CIPS-IPS at ambient temperature and pressure (bottom trace), at a temperature just above $T_C$ (350 K, middle trace), and at a pressure of 2.5 GPa (top trace). Two major differences can be seen between the room temperature spectrum and the spectrum collected just above $T_C$ (i.e. bottom and middle traces in Fig. 4), both of which are denoted by red dotted vertical lines in Fig. 3 – the loss of intensity of a peak around 100 cm$^{-1}$, and a concomitant increase in intensity of a peak around 550 cm$^{-1}$. As mentioned above, the peaks at 100 and 550 cm$^{-1}$ are both attributed to anion librations and stretching modes in the CIPS sub-lattice in CIPS-IPS. As the crystal goes through the paraelectric transition, both cations undergo displacements such that the In$^{3+}$ ion moves into a more central position within the S$_6$ octahedral cage, whereas the Cu$^+$ ions equally occupy both the up and down positions within its octahedron. The motion of the Cu$^+$ ions affects the S$_6$ octahedral cage, which in turn undergoes deformations as it accommodates the new cation distribution. These octahedral deformations are likely the cause for the decreased (increased) intensity of the 100 (550 cm$^{-1}$) peak above $T_C$. Note that pure-phase IPS is consistently paraelectric at all temperatures and hence we do not see any changes in the peaks associated with IPS (127, 140, 580 and 610 cm$^{-1}$) as the temperature increases.

Returning to the CIPS-IPS crystal, comparing the spectrum at 350 K with the spectrum at 2.5 GPa (i.e. the middle and top traces in Fig. 3), we do not observe a decrease (increase) in intensity of the 100 (550 cm$^{-1}$) peak, confirming that we do not see a ferrielectric-paraelectric transition in CIPS-IPS owing to the pressure-induced increase in $T_C$. However, we do observe a marked reduction in the intensity of Raman peaks at 127 and 255 cm$^{-1}$ (indicated in Fig. 4 by the blue dashed lines). Note that the corresponding peak frequencies at 2.5 GPa are blueshifted from their ambient pressure values, and appear at 133.5 and 261 cm$^{-1}$. As mentioned above, both the 127 and 255 cm$^{-1}$ peaks are assigned to anion vibrations in the IPS sub-lattice, and a reduction of intensities of these peaks gives us a vital clue into structural changes upon compression. Under ambient conditions, the IPS sub-lattice exerts a chemical pressure on the CIPS domains, and our above x-ray diffraction measurements of the layer spacings (Fig. 1c) indicate that the heterostructure formation compresses CIPS and expands IPS both along the stacking direction and within layers. Thus, it is reasonable to expect the already expanded IPS to undergo structural distortions when pressure is first applied. The decrease in intensity of the 127 cm$^{-1}$ peak, which corresponds to long-range anion librations, suggests an extended range migration and hopping of the In$^{3+}$ ions among their vacant sites, accompanied by a reduction in the IPS domains within each layer.

All of these structural changes are also visible in the optical microscope during compression. In Fig. 2c, we show optical microscope images taken from a CIPS-IPS crystal inside the DAC during compression. At ambient conditions, a pale yellowish crystal is clearly seen in the center of the image, and its color progressively darkens upon compression. The darkening increases up to ~5 GPa, above which nothing is visible beyond 5 GPa, and this persists up to the highest pressure (20 GPa). The change in color or reflectance of the crystal under pressure is a direct consequence of piezochromism and the resulting change in the electronic structure of the material. This effect has been observed previously in other metal thiophosphates [22,36] and attributed to lowering of the electronic bandgap and subsequent metallization, and is discussed further below. Upon decompression, the crystal regains its original color, supporting our hypothesis of recovery of the layered structure when pressure is released.

To highlight all the changes in the phonon modes in CIPS-IPS throughout our pressure range (*i.e.* up to 20 GPa), a 2D heat map of the Raman spectra (normalized with respect to the diamond peak intensity) is plotted in Fig. 5. In addition to the structural changes in the IPS sub-lattice below 2 GPa, we observe distinct discontinuities at four other pressures; these are indicated by the horizontal dashed lines in Fig. 5. Between 2 and 10 GPa, we observe a severe reduction of intensities of the low-frequency anion librational modes (< 200 cm$^{-1}$). The peaks that remain visible exhibit a monotonous hardening of their frequencies with increasing pressure; this hardening persists throughout our measured pressure range. We do not observe softening in any modes, indicating a lack of soft modes. This also suggests a lack of a displacive phase transition in CIPS-IPS and therefore any phase transition is likely to be of the order-disorder type (similar to that observed in CIPS). Above 10 GPa, we see an abrupt increase in peak intensities. These increases in intensity are seen much more clearly in Fig. 5 compared to Fig. 3 due to normalization of the spectra with respect to the Raman peak intensity from the diamond cell.

To better understand and correlate the spectral evolution to structural or electronic changes upon compression, we plot the frequencies, normalized intensities and FWHM (full width at half-maximum intensity) against pressure for seven peaks from the CIPS-IPS Raman spectrum (Figs. 6a-d). These peaks correspond to anion libration ($P_8$, 139.4 cm$^{-1}$ at ambient pressure, Fig. 6a), anion deformation ($P_{16}$, $P_{17}$ and $P_{18}$, at 270, 283.5 and 299.4 cm$^{-1}$, respectively, Fig. 6b), P-P stretching ($P_{20}$, 378 cm$^{-1}$, Fig. 6c) and anion stretching ($P_{23}$ and $P_{25}$, at 560 and 581.5 cm$^{-1}$, respectively, Fig. 6d).

Under compression, pure-phase CIPS is known to undergo a monoclinic-trigonal structural phase transition around 5 GPa [14]. This transition is accompanied by several changes in the Raman spectra, notably, a reduction in the number of peaks owing to the higher symmetry of the trigonal crystal structure, as well as the appearance of new peaks between 200 and 350 cm$^{-1}$. But unlike the case of CIPS, we do not observe any new peaks appearing around 5 GPa. We do observe a sharpening of the P-P stretching vibration ($P_{20}$, Fig. 6c), as well as the two anion stretching modes ($P_{23}$ and $P_{25}$, Fig. 6d) up to 5 GPa. In fact, the greatest changes in the spectra are observed around 10 GPa with a drastic reduction in the number of peaks and a sharp increase in intensity of the remaining peaks. The increase in intensity can be seen clearly in the heat map (Fig. 5) as well as Figs. 6a-d. The decrease in the number of peaks is akin to what is observed for the monoclinic-trigonal phase transition in CIPS around 5 GPa [14]. We thus propose that CIPS-IPS also undergoes a transition from the monoclinic phase to a trigonal crystal structure, but at a higher pressure ~ 10 GPa. It is not clear from the spectra how the CIPS and IPS sub-lattices rearrange through this phase transition. However, the evolution of the low frequency peaks provides insights. Throughout our compression study, we see that the Raman peaks between 100 – 300 cm$^1$ (anion librations) do not disappear completely (a higher magnification view of this region is shown in Fig. S2 [33]). Moreover, a peak around 88 cm$^{-1}$, arising from cation vibrations, persists throughout our measured pressure range. These trends suggest that the cation diffusion, which begins below 2 GPa, continues through the structural transition at 10 GPa. The cation hopping likely results in a reduction and redistribution of the CIPS and IPS sub-lattices, eventually leading to a homogeneous phase at high pressures. This is similar to the structure observed at high temperatures [11].

The overall increase in peak intensities above 10 GPa (Figs 4,5 and 6) could be attributed to the higher crystal symmetry of the high pressure phase. However, the monotonic increase in intensity of $P_{20}$, $P_{23}$ and $P_{25}$ between 10 – 17 GPa (Figs. 6c and 6d) hints at a different mechanism. One possibility is the resonance-enhanced increase in intensities owing to bandgap reduction with pressure. The electronic bandgap of MTP materials is known to decrease upon compression, with observations of indirect-to-direct bandgap transition in $SnP_2S_6$ [18], semiconductor-metal transitions in $CrPS_4$ [20], $MnPS_3$ [22] and $NiPS_3$ [37], and even superconductivity at high pressures and low temperatures in $FePSe_3$ [21,38]. The bandgap of CIPS under ambient conditions is ~2.8 eV [39], and a decrease the bandgap on compression would bring it closer in resonance to our Raman excitation energy (1.96 eV). A pressure-induced change in the electronic structure of CIPS-IPS is further supported by an anomalous peak sharpening for some of the peaks ($P_8$, Fig. 6a and $P_{23}$, Fig.6d) seen between 13 and 17 GPa. Owing to anharmonic effects, Raman modes typically harden and broaden with increasing pressure [40]. While we do observe the hardening of all the modes over the entire pressure range, the decrease in the FWHM of peaks between 13 to 17 GPa without affecting peak frequencies suggests an isostructural electronic transition. Such transitions, accompanied by metallization, have been observed previously in other ferroelectric materials such as $Hg_3Te_2X_2$ (X = Cl, Br) [41], $PbCrO_3$ [42] and $AgBiSe_2$ [43]. Pressure-driven modulation of the electronic structure is also supported by our observation of color changes at higher pressures (Fig. 1c).

At the high pressure limit of our study, we observe a plateau or a decrease in peak frequencies, accompanied by a decrease in intensities and broadening of the peaks (Fig. 6). This occurs between 16 and 17 GPa and is possibly due to an onset of metallization of the CIPS-IPS at high pressures, similar to what has been observed in other MTP materials [44]. Pressure-dependent electrical resistance measurements are needed to verify our hypothesis, in addition to multi-excitation Raman spectroscopy studies under pressure for verification of resonance-enhanced peak intensities. Future computational studies can also address the question of bandgap reduction and metallization.

Finally, we investigate the CIPS-IPS structure following decompression. Fig. 7a shows Raman spectra collected from the as-prepared sample prior to compression (labeled before) and after decompression and return to ambient pressure (labeled after). The spectra are normalized with respect to the highest intensity peak around 255 cm$^{-1}$. A number of differences in the intensities of peaks can be observed in the spectrum after decompression. Note that these two spectra are not collected from the same spot owing to sample movement under compression/decompression. However, spectra collected from multiple spots after decompression exhibit similar differences compared to the spectrum before compression. A few of the notable changes are decreases in intensities of peaks around 100 cm$^{-1}$ (anion librations in CIPS), 320 cm$^{-1}$ (anion deformation in CIPS) and 568 cm$^{-1}$ (anion stretching mode in CIPS). All three peaks correspond to vibrations in the CIPS sub-lattice, suggesting a significant structural rearrangement of the CIPS sub-lattice upon decompression. The corresponding anion libration modes from IPS, at 127 and 140 cm$^{-1}$, do not exhibit a significant reduction in their intensities. Likewise, the anion deformation mode around 255 cm$^{-1}$ retains its intensity after decompression, with a slight redshift in frequency (by 1 cm$^{-1}$), which could be attributed to residual tensile strain. The reduction of intensities of the CIPS Raman peaks also hints at the disappearance of ferroelectric order. In fact, the reduction of intensities of the same peaks is also observed upon rapid quenching from 510 K (i.e. above the structural phase transition temperature) to room temperature [11].

We reveal these structural changes through different microscopic characterization techniques. Figs. 7b-7e show the chemical composition of the CIPS-IPS crystal and the changes brought on through pressure. The secondary electron (SE) image (Fig. 7b) shows a smooth crystal surface without any evidence of multiple domains, even though we expect to observe two distinct domains corresponding to CIPS and IPS. However, these domains are revealed in the chemical maps obtained by electron dispersive spectroscopy (EDS). Fig. 7c shows the distinct chemical domains in the as-grown crystal. The In$_{4/3}$P$_2$S$_6$ domains (green in the image) manifest themselves as distorted diamonds in a CuInP$_2$S$_6$ matrix (red). The domains vary in size but are roughly 5-15 μm in length and 2-4 μm in width. When pressurized to ~20 GPa and released, the CIPS-IPS crystal undergoes a phase transition into a uniform single phase, as mentioned above. Upon the release of pressure, this single phase re-segregates into distinct CIPS and IPS phases, as visible in the Raman results. However, the domains are much smaller and cannot be resolved down to 1 μm resolution, roughly the ultimate resolution of our EDS detector (Fig. 7d and 7e).

Figs. 7f and 7g show AFM phase contrast images collected before and after compression. The corresponding topography images are included in the Supplemental Information, Fig. S3 [33]. The diamond-like IPS sub-lattices can be seen clearly in the AFM image from the as-grown crystal, with lateral sizes commensurate with the sizes seen in the EDS images. Some debris arising from the cleavage of the crystal can also be seen. The phase images of the decompressed CIPS-IPS sample (Fig. 7g) shows a complete loss of phase structure. We see some small striations which can potentially be attributed to size-reduced domains of the IPS sub-lattice. However, upon observation using a smaller scan window (~ 200 x 200 nm [33]), no structural information can be obtained. The transition to a crystalline yet chemically disordered (with respect to the metal cation) phase structure as evidenced by both EDS and AFM further suggests the loss of ferroelectric order and demonstrates the use of pressure to control the phase behavior and thus, the ferroelectric properties of the material.

## 4. Conclusions

We measured Raman spectra of Cu-deficient $CuInP_2S_6$ (CIPS-IPS) under hydrostatic pressures up to 20 GPa. By comparing the spectrum of CIPS-IPS to pure phase CIPS and IPS, we identify several vibrational modes unique to either sub-lattice. A number of changes to the phonon modes are observable upon compression and a detailed analysis of the frequencies, intensities and linewidths reveal four pressures at which a transition occurs. At the lowest transition pressure (~2 GPa), we observe significant reduction of the IPS peaks, suggesting that the structural transition is initiated by diffusion of In$^{3+}$ ions. This results in a phase transition around 10 GPa, which we attribute to a monoclinic-trigonal phase transition owing to the reduction in number of peaks in the higher symmetry crystal structure. We also observe anomalous sharpening of Raman peaks and a loss of intensity at the highest pressures, which are tentatively attributed to electronic transitions (bandgap reduction and metallization). Finally, compression results in a loss of ferroelectric order, which is confirmed by the reduction in the average CIPS and IPS domain size below the ferroelectric limit (~ 50 nm). By uncovering rich pressure-dependent physics in CIPS-IPS, we highlight hydrostatic pressure as a way to control the electronic and ferroelectic properties in this novel 2D material.

## Acknowledgements

We acknowledge support through the United States Air Force Office of Scientific Research (AFOSR) LRIRs 19RXCOR052 and 18RQCOR100, AOARD-MOST Grant Number F4GGA21207H002 and the National Research Council Postdoctoral Fellowship award. Use of the Advanced Photon Source, an Office of Science User Facility operated for the U.S. Department of Energy (DOE) Office of Science by Argonne National Laboratory, was supported by the U.S. DOE under Contract No. DE-AC02-06CH11357.

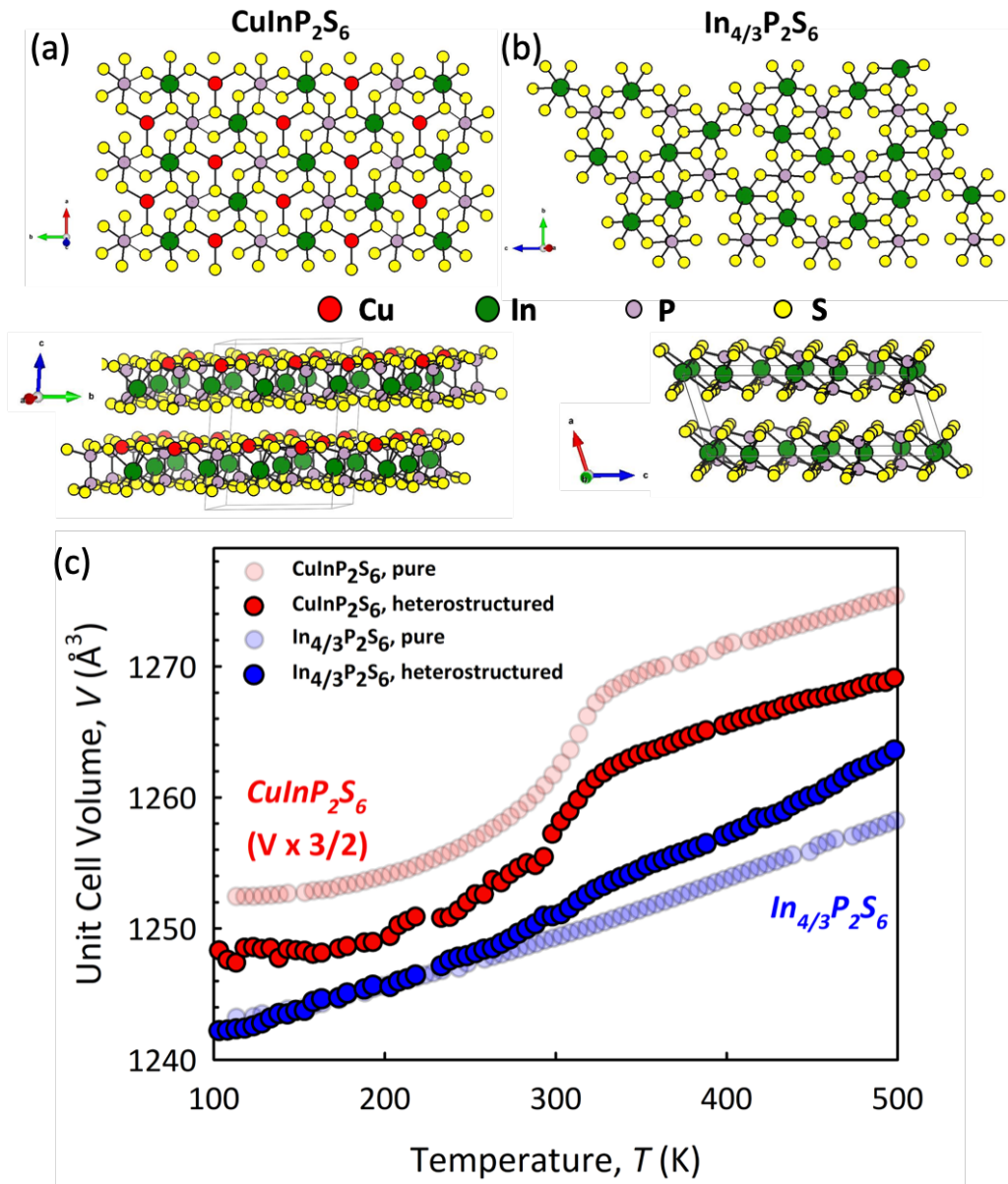

FIG. 1. Top and lateral views of the (a) CIPS and (b) IPS crystal structures. (c) Refinements of synchrotron XRD data showing the temperature-dependent evolution of the unit cell dimensions of pure-phase CIPS, pure-phase IPS, and the two respective phases in the CIPS-IPS in-plane heterostructure.

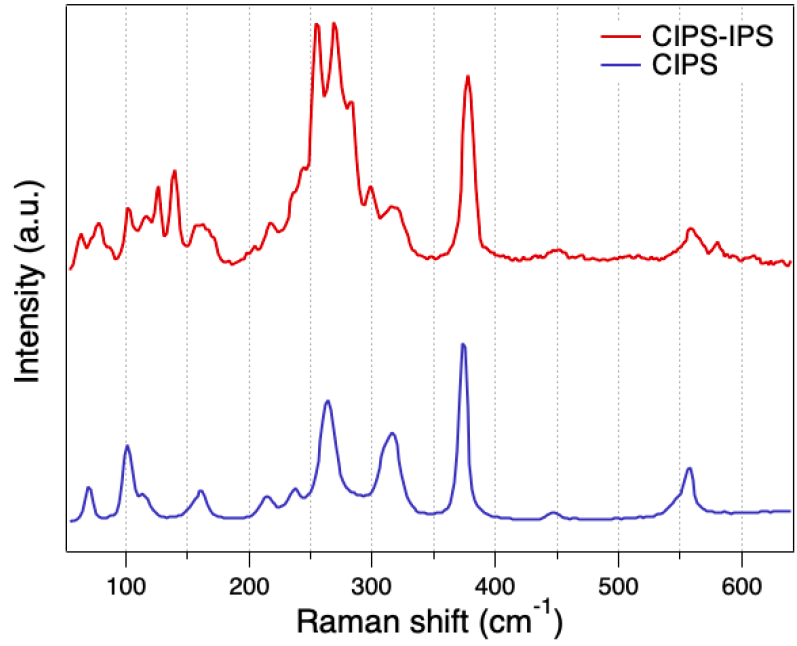

FIG.2. Room temperature ambient pressure Raman spectra (633 nm excitation) from CIPS-IPS and CIPS.

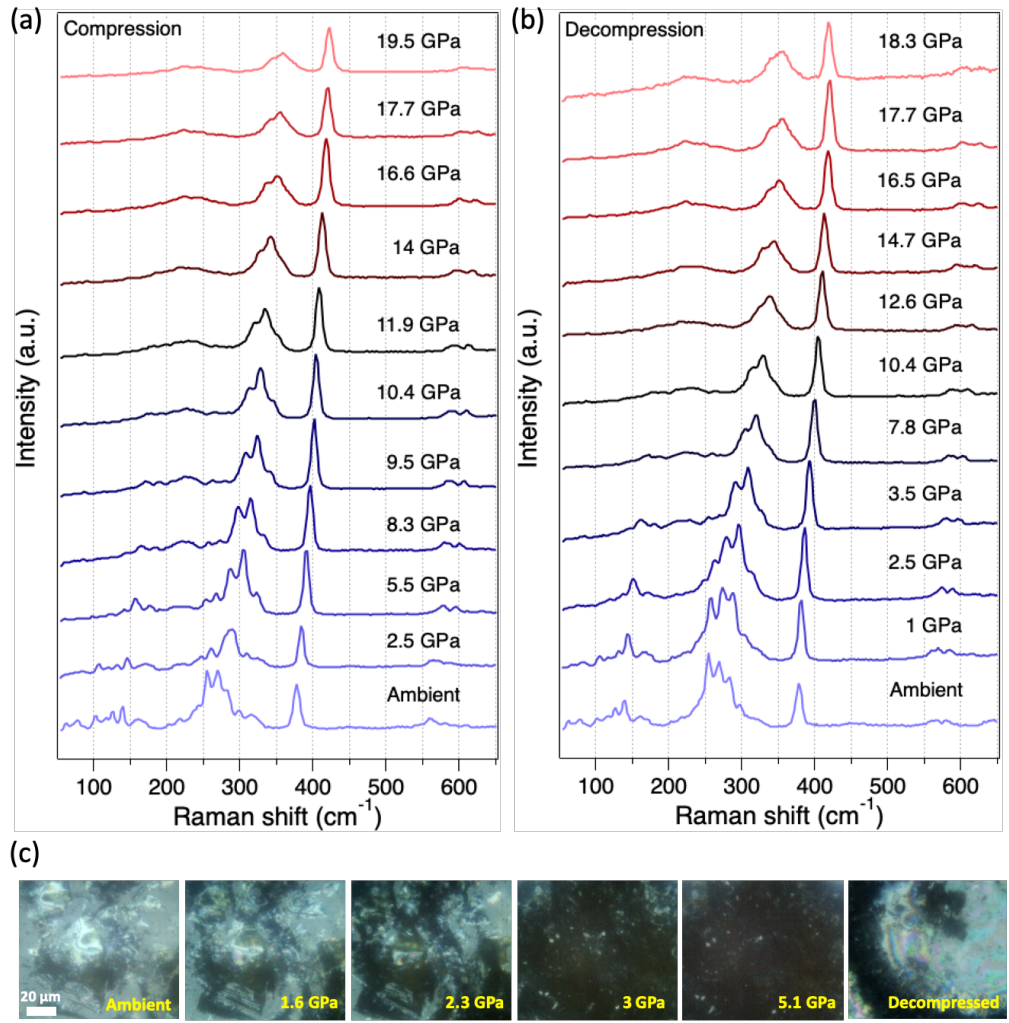

FIG. 3. Raman spectra from CIPS-IPS single crystals upon (a) compression and (b) decompression. All spectra are normalized with respect to the highest intensity peak and vertically offset for clarity. (c) Optical microscope images of the CIPS-IPS crystals inside the diamond anvil cell during compression up to 5.1 GPa. The color change indicates a piezochromic effect due to a change in the electronic structure.

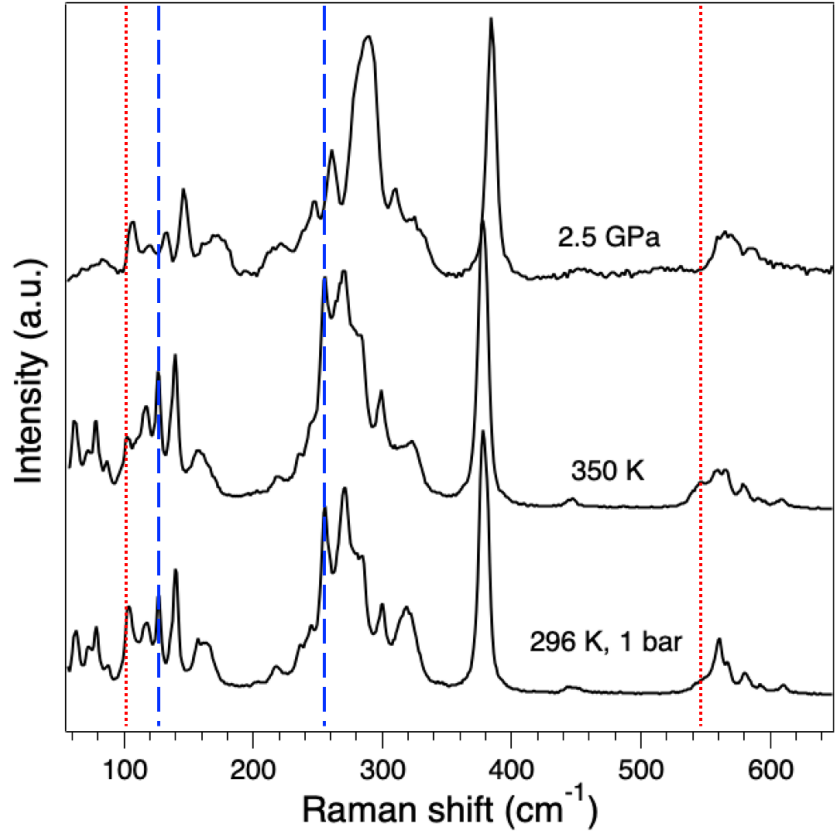

FIG. 4. Comparison between Raman spectra from CIPS-IPS at low pressure (and room temperature) and across the ferrielectric-paraelectric transition temperature ($T_C$ ~ 340 K). The dotted (dashed) lines indicate peaks with discernable changes in intensities upon heating (compression).

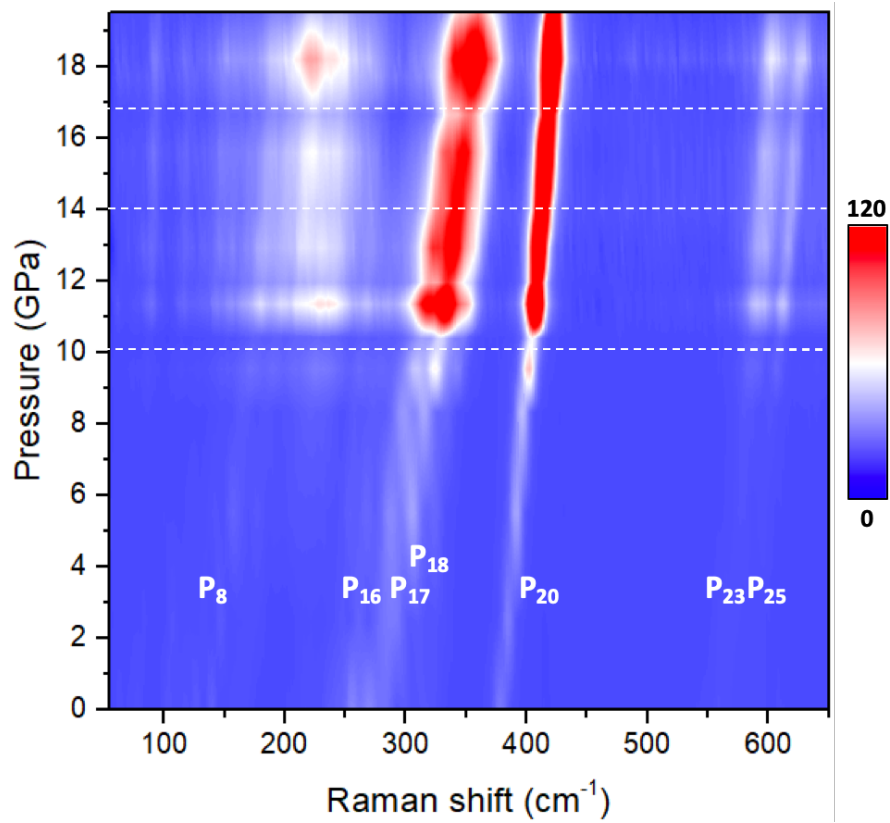

FIG. 5. 2D heat map showing the shift in Raman peak frequencies from CIPS-IPS upon compression. The color scale corresponds to peak intensities (counts) normalized to the diamond Raman peak.

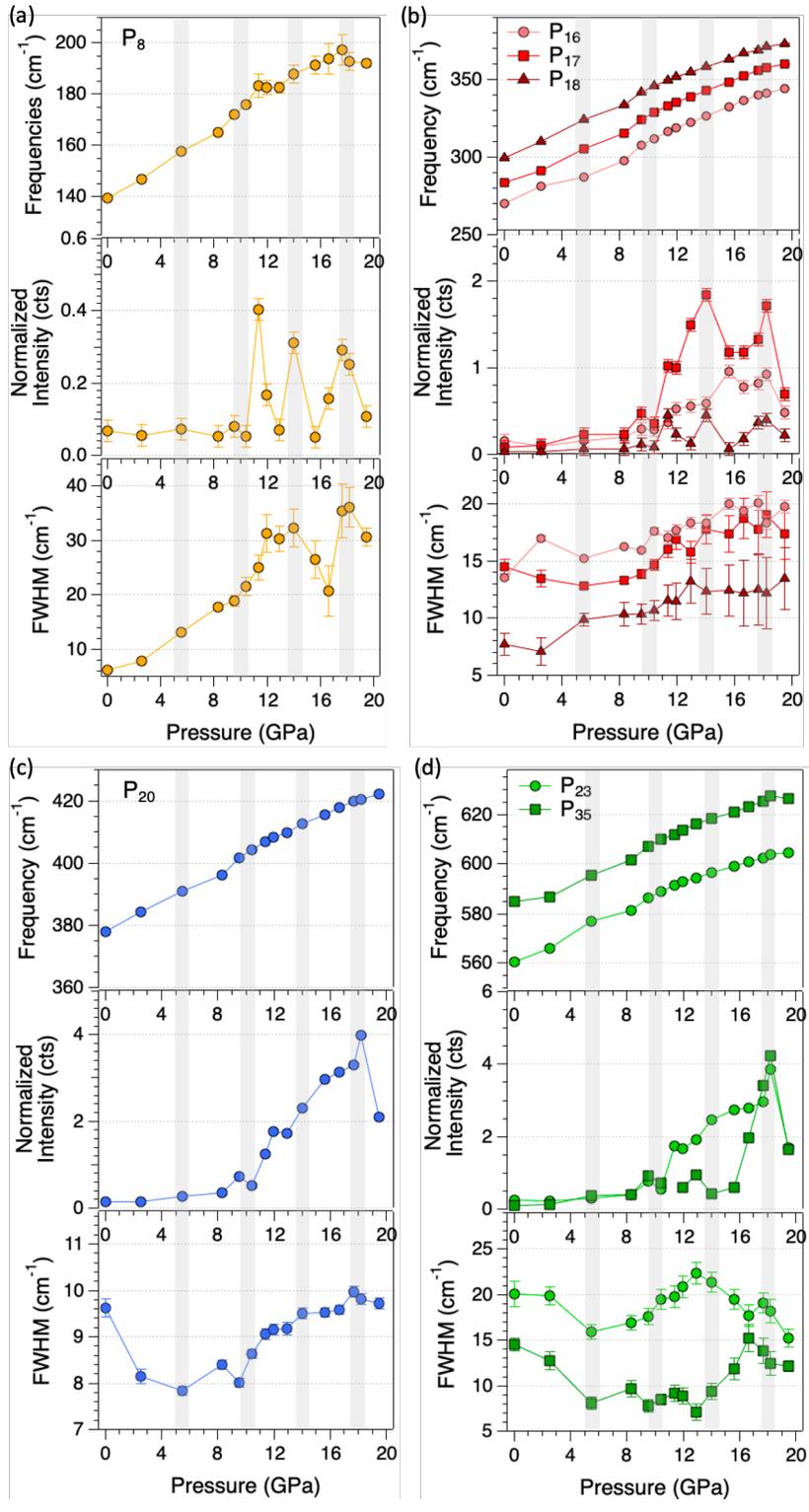

FIG. 6. Evolution of peak frequencies, normalized intensities and linewidths with pressure.

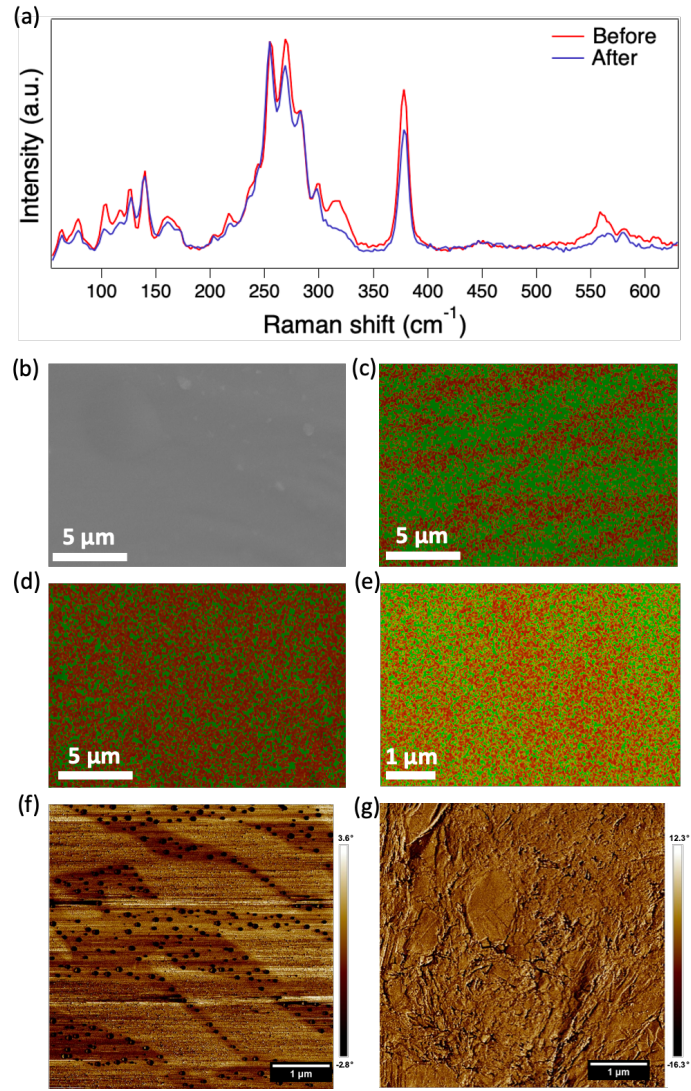

FIG 7. Structural analysis before and after compression. (a) Raman spectrum collected before compression and after decompression back to ambient pressure, and EDS (or AFM) maps showing the evolution of the microstructure in the as-prepared single crystals and after decompression. (b) Secondary electron image of the CIPS-IPS heterostructured crystal showing smooth surface. The striations in the EDS image of the as-grown CIPS-IPS crystal (c) correspond to the CIPS (red) and IPS (green) sub-lattices. These are significantly reduced in size upon decompression (d and e), leading to a loss of ferroelectric order. (f) AFM phase image from a freshly-cleaved surface of the as-prepared CIPS-IPS crystal. The domains of the CIPS and IPS can be clearly distinguished. (g) AFM phase image of the decompressed crystal showing a loss of domain structure.

Table 1. Raman peak frequencies from CIPS-IPS under ambient conditions and their peak assignments

| Peak | Frequency (cm⁻¹) | Assignment |
|---|---|---|
| 1 | 62.7 | Cation |
| 2 | 71.4 | Cation |
| 3 | 78.2 | Cation |
| 4 | 87.8 | Cation |
| 5 | 103 | Anion librations |
| 6 | 116.4 | Anion librations |
| 7 | 126.6 | Anion librations |
| **8** | **139.4** | δ(S-P-P) |
| 9 | 159.5 | δ(S-P-P) |
| 10 | 170.7 | δ(S-P-P) |
| 11 | 203 | δ(S-P-S) |
| 12 | 217.3 | δ(S-P-S) |
| 13 | 236.4 | δ(S-P-S) |
| 14 | 243.6 | δ(S-P-S) |
| 15 | 255.3 | δ(S-P-S) |
| **16** | **269.8** | δ(S-P-S) |
| **17** | **283.5** | δ(S-P-S) |
| **18** | **299.4** | δ(S-P-S) |
| 19 | 311.8 | δ(S-P-S) |
| **20** | **378** | ν(P-P) |
| 21 | 448.4 | ν(P-S) |
| 22 | 548.5 | ν(P-S) |
| **23** | **560.1** | ν(P-S) |
| 24 | 567.7 | ν(P-S) |
| **25** | **581.5** | |
| 26 | 594.8 | } $\nu$(P-S) |
| 27 | 609.5 | |

# References

[1] J. R. Schaibley, H. Yu, G. Clark, P. Rivera, J. S. Ross, K. L. Seyler, W. Yao, and X. Xu, *Valleytronics in 2D Materials*, Nat. Rev. Mater. **1**, 1 (2016).

[2] X. Liu and M. C. Hersam, *2D Materials for Quantum Information Science*, Nat. Rev. Mater. **4**, 669 (2019).

[3] N. R. Glavin, R. Rao, V. Varshney, E. Bianco, A. Apte, A. Roy, E. Ringe, and P. M. Ajayan, *Emerging Applications of Elemental 2D Materials*, Adv. Mater. **32**, 1904302 (2020).

[4] Z. Guan, H. Hu, X. Shen, P. Xiang, N. Zhong, J. Chu, and C. Duan, *Recent Progress in Two-Dimensional Ferroelectric Materials*, Adv. Electron. Mater. **6**, 1900818 (2020).

[5] C. Gong and X. Zhang, *Two-Dimensional Magnetic Crystals and Emergent Heterostructure Devices*, Science **363**, (2019).

[6] M. A. Susner, M. Chyasnavichyus, M. A. McGuire, P. Ganesh, and P. Maksymovych, *Metal Thio- and Selenophosphates as Multifunctional van Der Waals Layered Materials*, Adv. Mater. **29**, 1602852 (2017).

[7] M. A. McGuire, *Cleavable Magnetic Materials from van Der Waals Layered Transition Metal Halides and Chalcogenides*, J. Appl. Phys. **128**, 110901 (2020).

[8] M. A. Susner, R. Rao, A. T. Pelton, M. V. McLeod, and B. Maruyama, *Temperature-Dependent Raman Scattering and x-Ray Diffraction Study of Phase Transitions in Layered Multiferroic $CuCrP_2S_6$*, Phys. Rev. Mater. **4**, (2020).

[9] V. Maisonneuve, V. B. Cajipe, A. Simon, R. Von Der Muhll, and J. Ravez, *Ferrielectric Ordering in Lamellar $CuInP_2S_6$.*, Phys. Rev. B Condens. Matter **56**, 10860 (1997).

[10] M. A. Susner, A. Belianinov, A. Y. Borisevich, Q. He, M. Chyasnavichyus, P. Ganesh, H. Demir, D. Sholl, D. L. Abernathy, M. A. McGuire, and P. Maksymovych, *High $T_C$ Layered Ferrielectric Crystals by Coherent Spinodal Decomposition*, ACS Nano **9**, 12365 (2015).

[11] M. A. Susner, M. Chyasnavichyus, A. A. Puretzky, Q. He, B. S. Conner, Y. Ren, D. A. Cullen, P. Ganesh, D. Shin, H. Demir, J. W. McMurray, A. Y. Borisevich, P. Maksymovych, and M. A. McGuire, *Cation–Eutectic Transition* via *Sublattice Melting in $CuInP_2S_6/In_{4/3}P_2S_6$ van Der Waals Layered Crystals*, ACS Nano **11**, 7060 (2017).

[12] M. Chyasnavichyus, M. A. Susner, A. V. Ievlev, E. A. Eliseev, S. V. Kalinin, N. Balke, A. N. Morozovska, M. A. McGuire, and P. Maksymovych, *Size-Effect in Layered Ferrielectric CuInP2S6*, Appl. Phys. Lett. **109**, 172901 (2016).

[13] J. A. Brehm, S. M. Neumayer, L. Tao, A. O'Hara, M. Chyasnavichus, M. A. Susner, M. A. McGuire, S. V. Kalinin, S. Jesse, P. Ganesh, S. T. Pantelides, P. Maksymovych, and N. Balke, *Tunable Quadruple-Well Ferroelectric van Der Waals Crystals*, Nat. Mater. **19**, 43 (2020).

[14] A. Grzechnik, V. B. Cajipe, C. Payen, and P. F. McMillan, *Pressure-Induced Phase Transition in Ferrielectric CuInP2S6.*, Solid State Commun. **108**, 43 (1998).

[15] P. Guranich, V. Shusta, E. Gerzanich, A. Slivka, I. Kuritsa, and O. Gomonnai, *Influence of Hydrostatic Pressure on the Dielectric Properties of $CuInP_2S_6$ and $CuInP_2Se_6$ Layered Crystals.*, J. Phys. Conf. Ser. **79**, 012009/1 (2007).

[16] P. P. Guranich, A. G. Slivka, V. S. Shusta, O. O. Gomonnai, and I. P. Prits, *Optical and Dielectric Properties of $CuInP_2S_6$ Layered Crystals at High Hydrostatic Pressure.*, J. Phys. Conf. Ser. **121**, 022015/1 (2008).

[17] V. S. Shusta, I. P. Prits, P. P. Guranich, E. I. Gerzanich, and A. G. Slivka, *Dielectric Properties of $CuInP_2S_6$ Crystals under High Pressure.*, Condens. Matter Phys. **49**, 91 (2007).

[18] K. Z. Rushchanskii, Y. M. Vysochanskii, V. B. Cajipe, and X. Bourdon, *Influence of Pressure on the Structural, Dynamical, and Electronic Properties of the $SnP_2S_6$ Layered Crystal*, Phys. Rev. B **73**, 115115 (2006).

[19] S. Rosenblum and R. Merlin, *Resonant Two-Magnon Raman Scattering at High Pressures in the Layered Antiferromagnetic NiPS$_3$*, Phys. Rev. B **59**, 6317 (1999).

[20] R. A. Susilo, B. G. Jang, J. Feng, Q. Du, Z. Yan, H. Dong, M. Yuan, C. Petrovic, J. H. Shim, D. Y. Kim, and B. Chen, *Band Gap Crossover and Insulator–Metal Transition in the Compressed Layered CrPS$_4$*, Npj Quantum Mater. **5**, 1 (2020).

[21] M. J. Coak, D. M. Jarvis, H. Hamidov, A. R. Wildes, J. A. M. Paddison, C. Liu, C. R. S. Haines, N. T. Dang, S. E. Kichanov, B. N. Savenko, S. Lee, M. Kratochvílová, S. Klotz, T. C. Hansen, D. P. Kozlenko, J.-G. Park, and S. S. Saxena, *Emergent Magnetic Phases in Pressure-Tuned van Der Waals Antiferromagnet FePS$_3$*, Phys. Rev. X **11**, 011024 (2021).

[22] N. C. Harms, H.-S. Kim, A. J. Clune, K. A. Smith, K. R. O'Neal, A. V. Haglund, D. G. Mandrus, Z. Liu, K. Haule, D. Vanderbilt, and J. L. Musfeldt, *Piezochromism in the Magnetic Chalcogenide MnPS$_3$*, Npj Quantum Mater. **5**, 1 (2020).

[23] L. Zhang, Y. Tang, A. R. Khan, M. M. Hasan, P. Wang, H. Yan, T. Yildirim, J. F. Torres, G. P. Neupane, and Y. Zhang, *2D Materials and Heterostructures at Extreme Pressure*, Adv. Sci. **7**, 2002697 (2020).

[24] M. A. Susner, M. Chyasnavichyus, A. A. Puretzky, Q. He, B. S. Conner, Y. Ren, D. A. Cullen, P. Ganesh, D. Shin, H. Demir, J. W. McMurray, A. Y. Borisevich, P. Maksymovych, and M. A. McGuire, *Cation–Eutectic Transition via Sublattice Melting in CuInP$_2$S$_6$/In$_{4/3}$P$_2$S$_6$ van Der Waals Layered Crystals*, ACS Nano **11**, 7060 (2017).

[25] C. Frontera and J. Rodriguez-Carvajal, *FULLPROF as a New Tool for Flipping Ratio Analysis: Further Improvements.*, Phys. B **350**, e731 (2004).

[26] S. Klotz, J.-C. Chervin, P. Munsch, and G. L. Marchand, *Hydrostatic Limits of 11 Pressure Transmitting Media*, J. Phys. Appl. Phys. **42**, 075413 (2009).

[27] H. K. Mao, J. Xu, and P. M. Bell, *Calibration of the Ruby Pressure Gauge to 800 Kbar under Quasi-Hydrostatic Conditions*, J. Geophys. Res. Solid Earth **91**, 4673 (1986).

[28] T. Babuka, K. Glukhov, Yu. Vysochanskii, and M. Makowska-Janusik, *Layered Ferrielectric Crystals CuInP2S(Se)6: A Study from the First Principles*, Phase Transit. **92**, 440 (2019).

[29] M. A. Susner, M. Chyasnavichyus, A. A. Puretzky, Q. He, B. S. Conner, Y. Ren, D. A. Cullen, P. Ganesh, D. Shin, H. Demir, J. W. McMurray, A. Y. Borisevich, P. Maksymovych, and M. A. McGuire, *Cation–Eutectic Transition via Sublattice Melting in CuInP$_2$S$_6$/In$_{4/3}$P$_2$S$_6$ van Der Waals Layered Crystals*, ACS Nano **11**, 7060 (2017).

[30] Yu. M. Vysochanskii, V. A. Stephanovich, A. A. Molnar, V. B. Cajipe, and X. Bourdon, *Raman Spectroscopy Study of the Ferrielectric-Paraelectric Transition in Layered CuInP$_2$S$_6$*, Phys. Rev. B **58**, 9119 (1998).

[31] Y. Mathey, R. Clement, J. P. Audiere, O. Poizat, and C. Sourisseau, *Structural, Vibrational and Conduction Properties of a New Class of Layer-Type MPS$_3$ Compounds: Mn$^{II}_{1−x}$M$^{I}_{2x}$PS$_3$ (M$^I$ = Cu, Ag)*, Solid State Ion. **9–10**, 459 (1983).

[32] O. Poizat, C. Sourisseau, and Y. Mathey, *Vibrational Study of Metal-Substituted MPS$_3$ Layered Compounds: M$^{II}_{1−x}$M$^{I}_{2x}$PS$_3$ with M$^{II}$= Mn, Cd, and M$^I$= Cu (X= 0.13) or Ag (X= 0.50): I. Comprehensive Infrared and Raman Analysis and Structural Properties*, J. Solid State Chem. **72**, 272 (1988).

[33] *See Supplemental Material at (Link) for a Fitted Raman Spectrum, Pressure-Dependent Raman Spectra in the Low-Frequency Range and Additional AFM Images (Topography and High-Resolution)*.

[34] C. Sourisseau, J. P. Forgerit, and Y. Mathey, *Vibrational Study of the [P$_2$S$^{4-}_6$] Anion, of Some MPS$_3$ Layered Compounds (M = Fe, Co, Ni, In$_{2/3}$), and of Their Intercalates with [Co($\eta^5$-C$_5$H$_5$)$_2^+$] Cations*, J. Solid State Chem. **49**, 134 (1983).

[35] A. N. Morozovska, E. A. Eliseev, S. V. Kalinin, Y. M. Vysochanskii, and P. Maksymovych, *Stress-Induced Phase Transitions in Nanoscale CuInP$_2$S$_6$*, Phys. Rev. B **104**, 054102 (2021).

[36] S. V. Ovsyannikov, N. V. Morozova, I. V. Korobeinikov, V. Haborets, R. Yevych, Y. Vysochanskii, and V. V. Shchennikov, *Tuning the Electronic and Vibrational Properties of Sn2P2Se6 and Pb2P2S6 Crystals and Their Metallization under High Pressure*, Dalton Trans. **46**, 4245 (2017).

[37] X. Ma, Y. Wang, Y. Yin, B. Yue, J. Dai, J. Ji, F. Jin, F. Hong, J.-T. Wang, Q. Zhang, and X. Yu, *Dimensional Crossover Tuned by Pressure in Layered Magnetic NiPS$_3$*, ArXiv200914051 Cond-Mat (2020).

[38] Y. Wang, J. Ying, Z. Zhou, J. Sun, T. Wen, Y. Zhou, N. Li, Q. Zhang, F. Han, Y. Xiao, P. Chow, W. Yang, V. V. Struzhkin, Y. Zhao, and H. Mao, *Emergent Superconductivity in an Iron-Based Honeycomb Lattice Initiated by Pressure-Driven Spin-Crossover*, Nat. Commun. **9**, 1914 (2018).

[39] I. P. Studenyak, V. V. Mitrovcij, G. S. Kovacs, M. I. Gurzan, O. A. Mykajlo, Y. M. Vysochanskii, and V. B. Cajipe, *Disordering Effect on Optical Absorption Processes in CuInP2S6 Layered Ferrielectrics*, Phys. Status Solidi B **236**, 678 (2003).

[40] G. Lucazeau, *Effect of Pressure and Temperature on Raman Spectra of Solids: Anharmonicity*, J. Raman Spectrosc. **34**, 478 (2003).

[41] W. Cai, J. He, H. Li, R. Zhang, D. Zhang, D. Y. Chung, T. Bhowmick, C. Wolverton, M. G. Kanatzidis, and S. Deemyad, *Pressure-Induced Ferroelectric-like Transition Creates a Polar Metal in Defect Antiperovskites Hg$_3$Te$_2$X$_2$ (X = Cl, Br)*, Nat. Commun. **12**, 1509 (2021).

[42] S. Wang, J. Zhu, Y. Zhang, X. Yu, J. Zhang, W. Wang, L. Bai, J. Qian, L. Yin, and N. S. Sullivan, *Unusual Mott Transition in Multiferroic PbCrO$_3$*, Proc. Natl. Acad. Sci. **112**, 15320 (2015).

[43] V. Rajaji, P. S. Malavi, S. S. R. K. C. Yamijala, Y. A. Sorb, U. Dutta, S. N. Guin, B. Joseph, S. K. Pati, S. Karmakar, K. Biswas, and C. Narayana, *Pressure Induced Structural, Electronic Topological, and Semiconductor to Metal Transition in AgBiSe$_2$*, Appl. Phys. Lett. **109**, 171903 (2016).

[44] S. V. Ovsyannikov, H. Gou, N. V. Morozova, I. Tyagur, Y. Tyagur, and V. V. Shchennikov, *Raman Spectroscopy of Ferroelectric Sn$_2$P$_2$S$_6$ under High Pressure up to 40 GPa: Phase Transitions and Metallization*, J. Appl. Phys. **113**, 013511 (2013).